# High contrast all-optical spectrally distributed switching of femtosecond pulses in soft glass dual-core optical fiber

MATTIA LONGOBUCCO,[1,2] IGNAS ASTRAUSKAS,[3] AUDRIUS PUGŽLYS,[3] ANDRIUS BALTUŠKA,[3] DARIUSZ PYSZ,[1] FRANTIŠEK UHEREK,[4] RYSZARD BUCZYŃSKI,[1,2] IGNÁC BUGÁR [1,4,*]

[1]*Department of Glass, Łukasiewicz Research Network - Institute of Microelectronics and Photonics, Wólczyńska 133, 01-919 Warsaw, Poland*
[2]*Faculty of Physics, University of Warsaw, Pasteura 5, PL–02–093Warszawa, Poland*
[3]*Photonics Institute, Vienna University of Technology, Gusshausstrasse 27-387, 1040 Vienna, Austria*
[4]*International Laser Centre, Ilkovičova 3, 841 04 Bratislava, Slovakia*

*Corresponding author: ignac.bugar@ilc.sk

**All-optical switching of 77 fs pulses centered at 1560 nm, driven by 270 fs, 1030 nm pulses in a dual-core optical fiber exhibiting high index contrast is presented. The fiber is made of a thermally matched pair of lead silicate and borosilicate glasses used as core and cladding material, respectively. The novel switching approach is based on nonlinear balancing of dual-core asymmetry, by control pulse intensity induced group velocity reduction of the fast fiber channel. Due to the fiber core made of soft glass with high nonlinearity high switching contrast exceeding 20 dB is attained under application of control pulses of only few nanojoule energy. The optimization of the fiber length brought the best results at 14 mm, which is in good correspondence with the calculated coupling length at the signal wavelength. The results express significant progress in comparison to similar studies based on self-switching of solitonic pulses in dual-core fibers and represent high application potential.**

## 1. Introduction

In information technology, all-optical signal processing is one of the targeted routes toward higher-capacity and higher-data-rate optical networks. It is also required for developing digital optical computers that are expected to significantly outperform their electronic counterparts in many respects. Beside various optical arithmetic and logic operations [1,2], a frequent signal manipulation task is the cross-connect switch. It is indispensable in the realization of optical buffers [3], wavelength division multiplexed interconnects [4] and data center networks [5]. Dual-core optical fibers represent one of the simplest architectures to ensure the all-optical cross-connect switching. They comprise two inherently parallel channels, the interaction between which is easily controllable both by nonlinear or phase sensitive excitation of one or both input channels. Unlike more sophisticated structures, such as those based on the use of metamaterials [6], ring resonators [7] or plasmonic waveguides [8], dual-core fibers are much simpler to produce. Moreover, all of the mentioned alternative switching devices were developed in a single channel form. Therefore, these devices have to be duplicated and outfitted with a proper input signal splitter to be able to act as a cross-connect switch, increasing the cost and complexity even further. One of the most elegant approaches to cross-switching is the nonlinear loop mirror (NOLM), which recently ensured high contrast dropping of signal from 10 Tb/s data rate stream requiring only control pulses at the level of 2 pJ [9]. It was demonstrated over a 300 km fiber link in a configuration, which ensure seamless integration into an optical fiber communication network. On the other hand, such an ultrafast device with a low power consumption comprises 20 m of optical fiber in the loop and a wavelength division multiplexer on both the input and the output ends required to separate the signal and control pulses. Consequently, the NOLM technique is rather bulky, complex and complicated to cascade, thus its application in a multi-channel cross-connect switch is highly questionable.

Therefore, dual core optical fibers (DCF) remain promising candidates for all-optical signal processing, particularly for cross-switching tasks [10,11]. A combination of a highly nonlinear glass material and micrometer-size fiber core diameters makes it possible to drastically lower the power consumption required for optical switching [12,13]. Moreover, device length at the level of 1 cm is easily attainable by this approach [11,13,14]. The opportunity to operate in the solitonic propagation regime ensures pulse shape preservation despite of the energy transfer between the coupled cores in the dual-core configuration, which was a subject of extensive theoretical groundwork [15-17]. Among other architectures, dual- and triple-core fibers were proposed as potential all-optical logic gates controlled with femtosecond solitonic pulses [18,19]. Due to the simplifications in numerical models and the stringent requirement for maintaining exact dual-core symmetry, it was not possible for a long time the experimental realization of the benefits predicted by theory [11,20]. Recently, we have succeeded in demonstration of a broadband switching in a dual-core fiber taking advantage of the solitonic propagation regime. We have developed a highly nonlinear, high index contrast DCF, which allowed self-switching of femtosecond solitons at 1700 nm [21] and also in the C-band [22] at sub-nanojoule switching pulse energies. In this paper, we present an approach of spectrally distributed control/signal switching in the same DCF, exhibiting slight dual-core asymmetry, which allows achieve even higher switching contrast and better pulse shape preservation.

## 2. Experimental conditions

The fiber fabrication procedure and the calculation of its basic optical properties were described in our previous work [21], now we introduce them here just shortly. The DCF was made using two thermally matched soft glasses PBG-08 (lead-silicate) and UV-710 (borosilicate) with index contrast at level of 0.4 in the near infrared. The highly nonlinear lead-silicate glass, with 20 times higher Kerr nonlinearity than the silica glass, was used as core material, while the borosilicate one as cladding material. The dispersion profile of the DCF was calculated considering only one separate fiber channel, which is the standard approach in the generally approved coupled mode theory [10] we used also in our earlier studies [14,20]. Based on these approach, it is possible to calculate the measure of the dual-core asymmetry, because the linear optical properties were acquired for the both fiber channels. In order to determine it, we used the real scanning electron microscope image of the fiber cross section (Fig.2) for the numerical simulation work. The discretization of the image allows the separate analysis of the both channels by a mode solver, considering them as a single core structure surrounded by homogeneous cladding of the low index glass. The numerical simulations brought effective refractive index difference $\Delta n = 0.00022$ [23] between the two channels in the C-band. While determining the coupling properties, the real dual-core structure was considered with some simplifications, which neglected its slight asymmetry [21]. In the frame of this paper we are focusing on the nonlinear balancing of the dual-core asymmetry, supposing equalized index of refraction at the proper peak power level of the control pulse. Therefore, the mentioned approach of the coupling length calculation neglecting the dual-core asymmetry is relevant in this case.

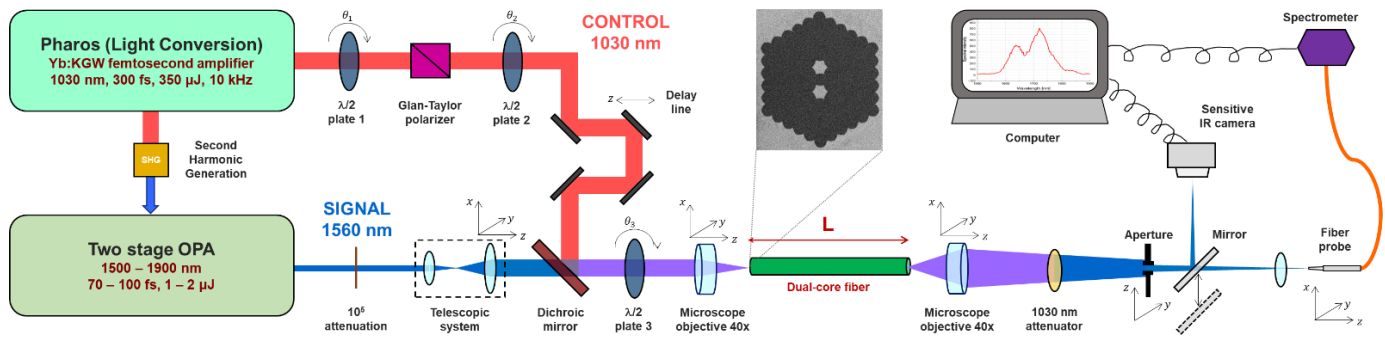

Fig.1. Experimental scheme containing the laser source of the synchronized control and signal pulses and the setups of the controlled dual-core fiber excitation and the output signal registration.

We launched two 10 kHz repetition rate sequences of timely synchronized femtosecond pulses with different central wavelength into one fiber core. The source of the 1030 nm control pulses was a commercial ultrafast Yb:KGW amplifier (Pharos, Light Conversion) giving pulses with width of 270 fs. The 1560 nm, 77 fs signal pulses were generated in a double pass OPA, pumped by frequency doubled output of the same Yb:KGW amplifier. The energy and the polarization of both sequences of pulses were independently controlled before their recombination by a dichroic mirror. The mirror directed the combined beam towards a microscopic objective (MO) with 40x magnification ensuring the in-coupling into one of the DCF cores. Delay line unit was used in the control arm to set the proper time shift between the signal and control pulses. Additionally, a telescopic system was placed into the signal arm with adjustable distance between their two objectives in order to set the same beam diameter of the both beams at the input fiber facet. The second MO imaged the output facet of the DCF onto an infrared camera chip (Xeva 1.7 320, Xenics) or onto a multimode collecting fiber attached to a spectrometer (NIRQuest, Ocean Optics). More details about the experimental setup ensuring femtosecond excitation and output field registration using identical DCF is presented in [21]. The energy of the signal pulses was set at 100 pJ level, which led to their linear propagation in the absence of control pulses. A series of camera images of the spatial distribution of the signal field were recorded with increasing the energy of control pulses in range of 1 – 10 nJ. The control field was suppressed behind the fiber by a dichroic mirror. Under unchanged experimental conditions, the spectra of signal pulses coming out from the excited and non-excited core were also separately measured. This whole procedure of registration of signal field dependence on control pulse energy was repeated for different fiber lengths in order to identify the optimal one for high contrast signal pulse switching.

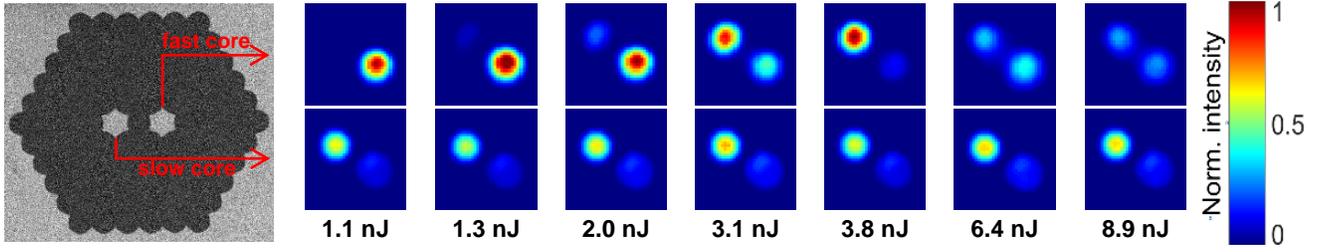

Fig.2. Scanning electron microscope image of the cross-section of the all-solid dual-core optical fiber structure (left). Infrared camera images of the 1560 nm, 77 fs signal field at the DCF output under increasing energy of 1030 nm, 270 fs control pulses, while the right (top series) and the left (bottom series) fiber core was excited by the combined beam.

## 3. Results and discussion

Fig.2 presents the dependence of the output signal field spatial distribution on the energy of the control pulses, while the right (top series) and the left (bottom series) fiber core was excited sequentially by the combined beam at optimized fiber length of 14 mm. The false colors in the images represent the normalized registered intensity on linear scale in accordance to the legend at the right side of Fig.2. The different behavior between the two series is obvious, because in the case of the bottom series any dependence of the dual-core extinction ratio on the control pulse energy was observed. In contrast, in the case of right core excitation, high-contrast switching took place. Only the right core was observed at initial 1.1 nJ control pulse energy; however, nearly the whole signal field was transferred to the opposite core at 3.8 nJ energy level. Further increase of the energy led again to the dominancy of the right core. The numerical simulation of the coupling length $L_C$ spectral profile, performed on the basis of the SEM image of the fiber cross-section, resulted in values of 13 mm and 116 mm at 1560 nm and 1030 nm wavelengths, respectively [21]. The significantly longer $L_C$ in the case of the control field in comparison to the fiber length (14 mm) causes propagation of the control pulses in the excited core with negligible energy transfer to the opposite one. The value of $L_C$ at the signal wavelength is very close to the experimentally determined optimal length. Both the shorter and the longer fibers resulted in worsening of the switching contrast. Therefore, the explanation of the efficient energy transfer at 3.8 nJ control pulse energy is the nonlinearly induced asymmetry balancing between the two fiber channels. The positive sign of the Kerr nonlinearity in the guiding glass causes decrease of the group velocity $v_g$ of the excited core in the time window of the control pulse duration to the level of the non-excited core $v_g$ [23]. Thus, the experimental results (Fig.2, top series) unveiled that the excited (right) core represent the fast and the non-excited one the slow propagation channel, respectively. Theoretical considerations predict increase of the energy transfer ratio between the cores with decreasing dual-core asymmetry up to the level 100 % in the case of totally symmetrical DCF [10]. This principle is responsible for the observed switching behavior at optimized fiber length. Under low control pulse energy excitation, the nonlinear interaction is negligible and the dual-core asymmetry prevents the energy transfer: the whole signal pulse energy remains in the excited core as it is observable on the camera image at 1.1 nJ. This camera image is identical to the registration made when the control beam is blocked. Then, the increase of the control pulse energy causes monotonic enhancement of the pulse energy transfer ratio to the non-excited core. Theory predicts that the nonlinear asymmetry balancing requires a specific pulse peak power [24]. In the case of both lower and higher power levels, just partial energy transfer can take place. In the case of lower pulse energies, the fast core remains still faster and above this level, it becomes slower in comparison to the non-excited (originally slow) core. The camera results exhibit this character because, above pulse energy level of 3.8 nJ, the excited (right) core output intensity decreases monotonically with increasing energy. In contrast to the lowest control pulse energy, it does not reach the zero level due to the strong nonlinear deformation of the signal field at higher control energies. However, camera pictures taken at high energy are important only for the explanation of the physical mechanism of the switching. For practical reasons, it is not necessary to use pulse energies above the optimal switching energy of 3.8 nJ.

In Fig.3 spectra registered at identical experimental conditions, i.e. right core excitation of the same DCF at 14 mm fiber length, are presented. Together with the spectral measurements, camera registrations in the same case of right core excitation were repeated in order to verify the results presented in top row of the Fig.2. Selected camera images are presented in Fig.3a, as insets. Fig.3a represents the dependence of the dual-core extinction ratio spectral profile $ER(\lambda) = 10\log(S_r(\lambda)/S_l(\lambda))$ on the control pulse energy. $S_{r,l}(\lambda)$ denote the spectral intensity in the right (excited) and left core, respectively. In this representation, the positive $ER(\lambda)$ values mean higher spectral intensity in the excited core, while the negative in the non-excited core. The dashed curve in Fig.3 shows the extinction ratio spectrum, which was acquired at 2.5 nJ pulse energy, situated closest to the 0 level, representing the best balance between curves $S_r(\lambda)$ and $S_l(\lambda)$ considering the whole spectral profile. The camera image registered at this energy reveals the same situation, similar level of intensities coming from the both cores. The general trend of the curves is a monotonic extinction ratio decrease within the whole spectral bandwidth with some anomalies at the wings. However, this trend is without exceptions in the spectral range of 1515 – 1565 nm, and the maximum change of the $ER$ (switching contrast) with the value of 21.4 dB was identified at the particular wavelength of 1560 nm. The switching contrast is evaluated as a difference in $ER$ when the control pulse energies are 1.1 and 3.8 nJ. Another advantageous aspect of the identified switching performance is the symmetrical displacement of the extreme $ER$ values with respect to the 0 level. Both the positive and the negative extremes of the dual-core extinction ratio are at the level of 10 dB. The two images of the output facet of the DCF, captured at the switching energies of 1.1 and 3.8 nJ shown in Fig.3a, confirm the clear dominancy of the appropriate core in the extreme cases. Moreover, the peak intensities in these images are rather similar, regardless which core is dominant. It reveals, that nearly the entire pulse energy was transferred between the two channels during the nonlinear switching process.

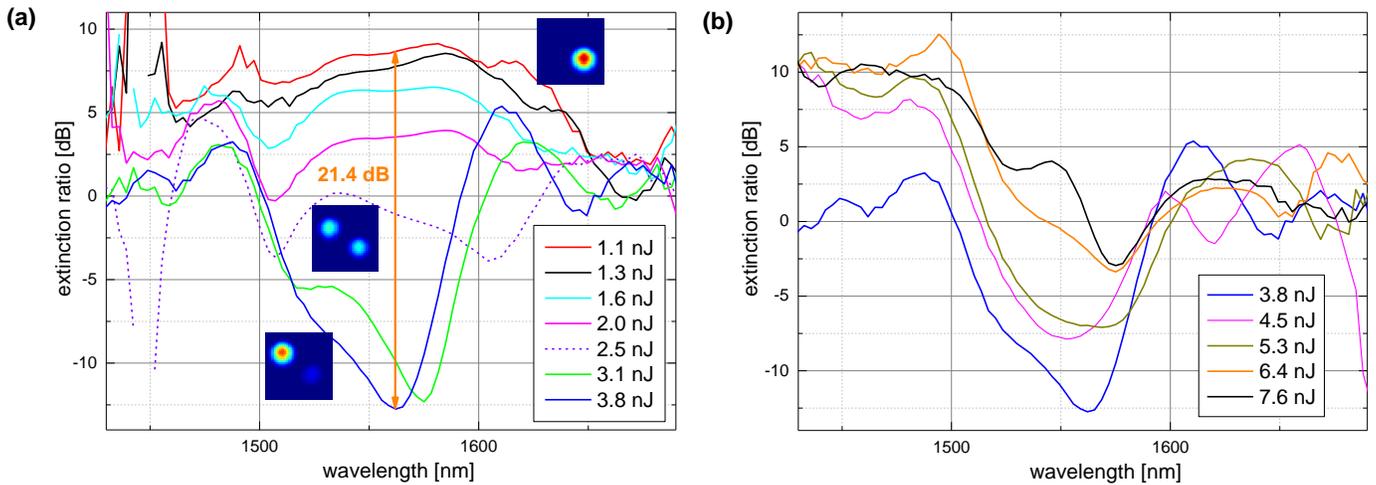

Fig.3. Dependence of extinction ratios spectral profile on control pulse energy (indicated) in range of **(a)** 1.1 – 3.8 nJ and **(b)** 3.8 – 7.6 nJ. Infrared camera images of the output fiber facet at energies 1.1, 2.5 and 3.8 nJ are in the insets, placed at the correspondent spectral curves.

Fig.3b presents the spectra of the extinction ratio recorded at control pulse energies ranging from 3.8 nJ to 7.6 nJ. The spectral profiles reflect an opposite trend in comparison to the lower energy case, i.e. the extinction ratio in the spectral range of 1515 – 1565 nm monotonically increases with the increasing control pulse energy. Thus, the two sets set of the measured spectra confirm the optimal switching energy of 3.8 nJ. Below this level decreasing and above it increasing trend of the $ER(\lambda)$ was observed which is in correspondence with the recorded camera images in Fig.2. It represents a further indication of the nonlinear balancing of the slight dual-core asymmetry, which ensures maximal transfer between the cores at a distinct control pulse energy. In Fig.4 normalized spectra of the output signal pulse collected always from the dominant core at different control pulse energies are presented. The lower energy spectra dominating in the excited core and depicted by solid lines have rather similar profiles, which indicates that the propagation of the signal pulses under these conditions is free from nonlinear deformations. At higher control pulse energies, the spectra dominating in the non-excited core are represented by dashed lines and exhibit different behavior. Namely, a monotonic blue shift and spectral narrowing tendency with increasing control pulse energy. The changes, however, are not significant: the central wavelength has shifted only by 5 nm as the bandwidth became narrower by 20% at the 3.8 nJ control pulse energy, in comparison to the low energy spectral profiles. Importantly, the spectra remain smooth and don't exhibit any asymmetry. The performed pulse diagnostics resulted in 77 fs pulse width, with negligible temporal dependence of the phase, which measurement was confirmed by the independent spectral registration giving 48 nm bandwidth at 1.1 nJ level. It corresponds to bandwidth limited pulse of 75 fs, therefore we can assume that the switching was accompanied just by a slight pulse stretching. Taking into consideration the initial pulse width, we estimate an increase to 92 fs at the DCF output. In the case of standard integral type registration used in optical communication technologies, the observed pulse lengthening should not have any pronounced effect on the switching performance. Moreover, such output signal is repeatedly switchable with high contrast by the same approach of the nonlinear balancing of dual-core asymmetry using the significantly longer control pulse.

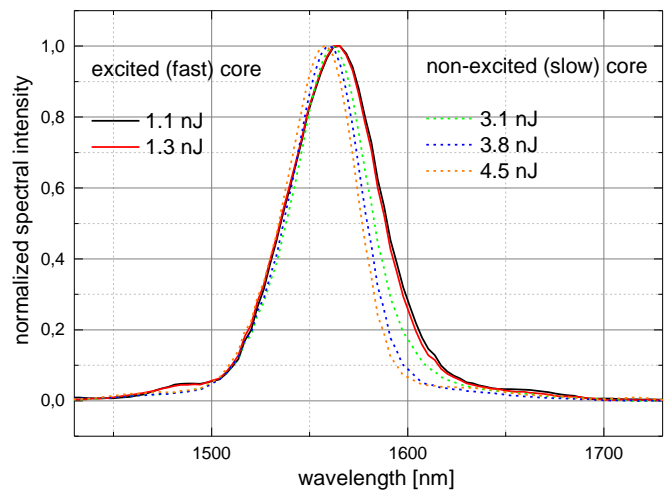

Fig.4. Normalized signal pulses spectra collected at low control pulse energies from the excited core (solid) and around the switching energy from the non-excited core (dashed).

It is worth mentioning that, as it follows from the camera images, the real switching contrasts are definitely higher than 21.4 dB. The extinction ratios were calculated also by acquisition of the output signal energy based on the series of the camera images. It was performed as separate integration of the field distribution in the vicinity of both cores, which procedure resulted in extinction ratio value for all applied control pulse energies $ER(E_{control})$ [21]. Such processing of camera images is a more general evaluation of the switching performance because brings information about the whole pulse energy in integral manner instead the spectrally resolved $ER(\lambda)$, presented in Fig.3. The evaluation of the whole series of $ER(E_{control})$ values at different control pulse energies resulted in maximal $ER$ change above 24 dB. It is worth mentioning, that this advantageous switching contrast was obtained without considering the lowest control pulse energies. It was caused by the observation, that at 1-1.1 nJ energy level the output field intensity was only at the level of the noise in the non-excited core, which field distribution character prevents the calculation of the extinction ratio. Therefore, the switching contrast should exceed even the 25 dB level, which is an excellent value from the application point of view. Unfortunately, this extraordinary performance is not obvious from the spectral results due to chromatic aberrations of the registration setup containing two lenses and iris aperture for proper isolation of the monitored core [20,21]. However, from the practical point of view, the camera registration is more relevant, because the majority of applications do not require spectral resolution of the output signal.

By taking into account the difference of effective refractive index of the two fiber cores, it is possible to estimate the control pulse energy, at which the balancing of the dual-core asymmetry can be achieved. The Kerr nonlinearity induces change of the refractive index $\Delta n = n_2 \cdot I$, where $n_2$ is the nonlinear index of refraction, which is equal to 4.3 $10^{-19}$ m$^2$/W in the case of the PBG-08 glass used for the fiber core and $I$ is the field intensity. Considering effective mode area of 1.86 µm$^2$, pulse duration of 270 fs and in-coupling efficiency of 50% [21] we obtain the value of 0.52 nJ for the balancing control pulse energy. It is approximately 7 times lower than the experimentally determined optimal switching energy of 3.8 nJ. Even though, it represents a good estimate taking into account the walk-off between the signal and control pulses. Considering the group velocity difference between wavelengths 1030 nm and 1550 nm, the walk-off was determined to be 152 fs for 14 mm long fiber. The walk-off between the pulses results in higher control pulse energy required to induce asymmetry balancing between the two fiber channels, since the signal pulse experiences lower control field intensities at its leading and trailing edges. During the experiment, the delay between the control and signal pulses was optimized for every change of the experimental conditions. We assume that the optimal switching was achieved in the situation when the control and signal pulses optimally overlap in the central part of the fiber. This means that at the two fiber ends the signal pulse was time shifted from the control pulse by approximately 76 fs. Another factor leading to higher experimentally determined control pulse energy enhancement is the attenuation and stretching of control pulse during propagation along the fiber in the normal dispersion region. Therefore, the estimated energy of 0.52 nJ supports the concept of nonlinear balancing of dual-core asymmetry. Since the walk-off can be minimized by choosing properly paired wavelengths of the control and switching pulses as well as by targeted redesigning of the fiber structure, a spectrally distributed switching with sub-nJ control pulses sounds feasible. Moreover, the reduction of the walk-off would allow the use of shorter control pulses, and, consequently, the switching at 100 pJ energy levels.

## 4. Conclusion

The above described spectrally distributed switching represents a novel trend in all-optical signal processing at ultrafast transfer rates requiring only a simple centimeter range dual-core optical fiber structure. Recently, several complex studies of such fibers as tools for cross-connect self-switching of solitonic femtosecond pulses in spectral region of optical communication systems were performed. However, the spectrally distributed switching based on nonlinear balancing of the DC asymmetry overcome recently demonstrated self-switching in numerous aspects. First of all, the identified switching contrast of more than 20 dB exceeds both, the best experimental results [21] and the theoretical predictions focusing on the ultrafast solitonic self-trapping mechanism in highly nonlinear DCFs [25]. Second, the spectrally distributed control/signal approach causes minimal pulse distortions, which results in spectrally homogeneous switching performance. It was never registered in the case of self-switching, which requires higher level of nonlinear interaction. Finally, those advantageous features were obtained at only 14 mm long fiber, which is shorter than 43 mm, where the highest contrast of self-switching was achieved at the same signal wavelength of 1560 nm [22]. A disadvantageous aspect of the spectrally distributed switching demonstrated in this work is the relatively high energy (few nanojoule) of control pulses. Obvious reasons for that is the relatively large control pulse duration, in comparison to the self-switching experiments, and the walk-off between the switching and control pulses. However, this drawback can be eliminated by improvement of the dual-core symmetry of the fiber and by better group velocity matching between the signal and control fields. The redesigning of the DCF structure has potential also for further reduction of the device length, because the coupling length depends strongly on the distance between the two cores. It would simultaneously eliminate also the unwanted walk-off effect. Furthermore, it is important to stress that there is no fundamental bottom limit regarding the signal pulse energy, which was at the level of 100 pJ during the presented experiments. The energy was set for the convenient registration of the output field by the available camera and spectrometer considering the 10 kHz repetition rate of the applied laser system. However, using more sensitive registration technique or higher repetition rates the signal pulses energy can be reduced even below 1 pJ. Such low energy operation meets already the recent standards of the optical communication systems. Finally, the presented low distortion of the signal pulses allows to use this cross-connect switching method in a cascaded scheme and this option increases the application potentials of DCFs. Thus, the demonstrated novel approach represents a promising direction in ultrafast all-optical signal processing with scaling potential in terms of control energy, fiber length and repeatability.

**Funding sources and acknowledgments.** This work was supported *by* by Slovak R&D Agency under the contracts No. APVV-17-0662 and SK-AT-2017-0026, by Austrian Agency OeAD under contract number SK 02_2018, by internal grant of Łukasiewicz Research Network - Institute of Electronic Materials Technology with No. S5-10-1032-20 and by EU under program H2020 ACTPHAST 4.0 (Grant No. 779472). I.A. A.P. and A.B. acknowledge financial support of the Austrian Research Promotion Agency FFG (Eurostars Eureka Project No. 12576 HABRIA; FFG Project No. 867822) and Austrian Science Fund FWF (P 27577-N27).

**Acknowledgments**. We acknowledge Evgeni Sorokin's lab space provided at TU Wien for the experimental work.